\documentclass[preprint,showpacs,preprintnumbers,amsmath,amssymb]{revtex4}

\usepackage{graphicx}% Include figure files
\usepackage{dcolumn}% Align table columns on decimal point
\usepackage{bm}% bold math

\begin{document}
\draft
\title{Near-field diffraction of fs and sub-fs pulses:
  super-resolutions of NSOM in space and time}
\author{S. V.  Kukhlevsky, M. Mechler, L. Csapo}
\address{Institute of Physics, University of Pecs, Ifjusag u. 6, Pecs 7624,
Hungary}
\author{K. Janssens}
\address{Department of Chemistry, University of Antwerp,
Universiteitsplein 1, B-2610 Antwerp, Belgium}
%\date{\today}
\begin{abstract}
The near-field diffraction of fs and sub-fs light pulses by
nm-size slit-type apertures and its implication for near-field
scanning optical microscopy (NSOM) is analyzed. The amplitude
distributions of the diffracted wave-packets having the central
wavelengths in the visible spectral region are found by using the
Neerhoff and Mur coupled integral equations, which are solved
numerically for each Fourier's component of the wave-packet. In
the case of fs pulses, the duration and transverse dimensions of
the diffracted pulse remain practically the same as that of the
input pulse. This demonstrates feasibility of the NSOM in which a
fs pulse is used to provide the fs temporal resolution together
with nm-scale spatial resolution. In the sub-fs domain, the
Fourier spectrum of the transmitted pulse experiences a
considerable narrowing that leads to the increase of the pulse
duration in a few times. This imposes a limit on the simultaneous
resolutions in time and space.
\end{abstract}
\maketitle

\section{Introduction}
The near-field diffraction of light by sub-wavelength apertures
and its implication for near-field scanning optical microscopy
(NSOM) has been investigated over the last decade~\cite{Poh,Nie}.
In NSOM, a sub-wavelength aperture illuminated by a continuous
wave is used as a near-field light source providing the super
(sub-wavelength) resolution in space. In the last few years, a big
interest of researchers attracted the study of near-field
diffraction of ultra-short light pulses aimed at obtaining the
simultaneous super-resolutions in space and time (for example, see
the studies~\cite{Lew,Betz,Xie,Amb,Nah,Bro,Mit1,Mit2,Kuk1} and
references therein). The studies of NSOM-pulse system primarily
dealt with the diffraction of relatively long (ps) far-infrared
pulses by $\mu$m-size apertures~\cite{Nah,Bro,Mit1,Mit2}. It was
shown that at particular experimental conditions a ps pulse
experiences significant spectral~\cite{Mit1,Mit2} and
temporal~\cite{Nah,Bro,Mit1,Mit2} deformations when diffracted by
a $\mu$m-size aperture. The deformations lead to a modification of
the temporal resolution associated with the incident ps-pulse. The
recent study \cite{Mit1} indicated that spatial resolution of the
NSOM performing with ps pulses is defined by the aperture size and
it is independent of the wavelength.

The use of ps pulses in NSOM provides ps temporal resolution
together with $\mu$m-scale spatial resolution. The achievement of
higher simultaneous resolutions in space and time is challenging.
The NSOM performing with a continuous wave provides nm-scale
spatial resolution~\cite{Poh,Nie}. High-harmonic generation
produces near 0.5-fs (500 as) wave-packets with ~50-nm central
wavelengths~\cite{Pau,Hen}. A method of generation of sub-as
pulses has been already suggested in Ref.~\cite{Kap}. The
perspective is the achievement of the simultaneous nm-scale
spatial resolution together with the fs or sub-fs temporal
resolution. In the visible spectral region, which is the most
important domain for potential applications, the NSOM-pulse system
can be realized using the conventional setup of NSOM.
Unfortunately, the capabilities and limits of simultaneous spatial
and temporal resolutions of such a system are not known. From a
theoretical point of view, the most important questions are the
degree of collimating, the duration of the ultra-short pulse past
the sub-wavelength aperture and the rate of spatial and temporal
broadening of the pulse farther from the aperture. To address
these questions, the near-field diffraction of fs and sub-fs
pulses by a nm-size slit-type aperture in a perfectly conducting
thick screen is theoretically studied in the present paper.

The article is organized as follows. The theoretical approach used
for finding the amplitude distribution of the ultra-short pulse in
the near-field diffraction zone of the sub-wavelength aperture is
described in Section II. Results of computer simulation of the
near-field diffraction and discussion of the capabilities and
limits of simultaneous spatial and temporal resolutions of the
NSOM-pulse system are presented in Section III.

\section{Theory}
Owing to the complicity of boundary conditions of the near-field
diffraction experiment, the solution of the Helmholtz
wave-equation even for a continuous wave can be obtained only by
extremely extended computations. In order to simplify the
computations, the NSOM model is usually restricted to two
dimensions of a slit-type aperture~\cite{Nee,Bet}. The combination
of NSOM with an ultra-short pulse makes the numerical analysis of
the problem even more difficult. In the present article, we
consider a model of the NSOM-pulse system based on the
transmission of fs and sub-fs pulses through a nm-size slit-type
aperture in a perfectly conducting thick screen. The amplitude
distribution of the diffracted wave-packet is found by using the
Neerhoff and Mur coupled integral equations~\cite{Nee,Bet}, which
are solved numerically for each Fourier's component of the
wave-packet.

Let us briefly describe the integral approach of Neerhoff and Mur,
which was developed in the studies~\cite{Nee,Bet} for finding the
near-field distributions of the amplitude and intensity of a
continuous wave diffracted by a sub-wavelength slit in a perfectly
conducting thick screen. In the region I, the continuous plane
wave falls onto the slit at an angle $\theta$ with respect to the
$z$-axis in the $x-z$ plane, as shown in Fig. 1. The slit width
and the screen thickness are $2a$ and $b$, respectively. The
magnetic field of the incident wave is assumed to be time harmonic
and both polarized and constant in the direction $y$:
\begin{eqnarray}
{\vec{H}}(x,y,z,t)=U(x,z){\exp}(-i\omega{t}){\vec{e}}_y.
\end{eqnarray}
The electric field of the incident wave is found by using the
Maxwell equations for the field ${\vec{H}}$. The restrictions in
Eq. 1 reduce the diffraction problem to one involving a single
scalar field in only two dimensions. The Green function
approach~\cite{Nee,Bet} uses the multipole expansion of the field
with the Hankel functions in the regions I and III and with the
waveguide eigenmodes in the region II. The expansion coefficients
are determined by using the standard boundary conditions for a
perfectly conducting screen. In region III, the near-field
distributions of the magnetic ${\vec{H}}(x,z,t)$ and electric
${\vec{E}}=E_{x}{\vec{e}}_{x}+E_{y}{\vec{e}}_{y}$ fields of the
diffracted wave are given by
\begin{eqnarray}
{\vec{H}}(x,z,t)=i{\frac{a}{N}}{\frac{\epsilon_3}{\epsilon_2}}{\sum_{j=1}^N}
H_0^{(1)}[k_{3}((x-x_j)^2+z^2)^{1/2}](D{\vec{U}}_0)_{j}{\exp}(-i\omega{t}){\vec{e}}_y,
\end{eqnarray}
\begin{eqnarray}
E_{x}(x,z,t)=-{\frac{a}{N}}{\frac{\sqrt{\epsilon_3}}{\epsilon_2}}{\sum_{j=1}^N}
{\frac{z}{((x-x_j)^2+z^2)^{1/2}}}H_1^{(1)}[k_{3}((x-x_j)^2+z^2)^{1/2}](D{\vec{U}}_0)_{j}
{\exp}(-i\omega{t}),
\end{eqnarray}
\begin{eqnarray}
E_{z}(x,z,t)=-{\frac{a}{N}}{\frac{\sqrt{\epsilon_3}}{\epsilon_2}}{\sum_{j=1}^N}
{\frac{x-x_j}{((x-x_j)^2+z^2)^{1/2}}}H_1^{(1)}[k_{3}((x-x_j)^2+z^2)^{1/2}]
(D{\vec{U}}_0)_{j}{\exp}(-i\omega{t}),
\end{eqnarray}
where $\epsilon_i$ and $k_i$ are respectively the permittivity and
the wave number in the regions $i$ = I, II and III;
$x_{j}=2a(j-1/2)/N-a$, with $j=1,2,...,N$ and $N>2a/z$;
$H_0^{(1)}$ and $H_1^{(1)}$ are the Hankel functions. The
coefficients $(D{\vec{U}}_0)_{j}$ are computed by solving a set of
the Neerhoff and Mur coupled integral equations~\cite{Bet}. For
more details of the integral approach of Neerhoff and Mur see
Refs.~\cite{Nee,Bet}.

We now consider the near-field diffraction of an ultra-short light
pulse. For the sake of simplicity, we assume that a pulse falls
normally ($\Theta=0$) on the aperture. The magnetic field of the
incident pulse is assumed to be Gaussian-shaped in time and both
polarized and constant in the direction $y$:
\begin{eqnarray}
{\vec{H}}(x,y,z,t)=U(x,z){\exp[-2\ln(2)(t/{\tau})^2]}{\exp}(-i\omega_{0}{t})
{\vec{e}}_y,
\end{eqnarray}
where $\tau$ is the pulse duration and $\omega_0$ is the central
frequency. The pulse can be composed in the wave-packet form of a
Fourier time expansion~\cite{Kuk2}:
\begin{eqnarray}
{\vec{H}}(x,y,z,t)=\int_{-{\infty}}^{{\infty}}{\vec{H}}(x,z,\omega)
{\exp}(-i\omega{t})d\omega.
\end{eqnarray}
The field distribution of the diffracted pulse is found by using
the expressions (2-4) for each Fourier's $\omega$-component of the
wave-packet (6). The algorithm was implemented numerically in
section III. In the computations, we used the discrete Fast
Fourier Transform (FFT) of the function ${\vec{H}}(x,y,z,t)$
instead of the integral composition (6). The spectral interval
$[\omega_{min},\omega_{max}]$ and the sampling points $\omega_i$
were optimized by matching the FFT composition to the original
function (5).
\section{Results and discussion}
In this section, the amplitude distribution of the diffracted
pulse is found by using the expressions (2-4) for each Fourier's
$\omega$-component of the wave-packet (6). In order to establish
guidelines for the computational results, we first consider the
dependence of the amplitude of a time-harmonic continuous plane
wave (a FFT $\omega$-component of the wave-packet) transmitted
through the aperture on the the wave frequency
$\omega=2{\pi}c/{\lambda}$. The amplitude of a transmitted FFT
$\omega$-component depends on the frequency $\omega$. Owing to
this effect, the Fourier spectra and duration of the wave-packet
are changed under propagation through the aperture leading to
modification of the temporal and spatial resolutions associated
with the input pulse. The dispersion for a time-harmonic
continuous wave is usually described by the normalized
transmission coefficient $T_{cw}$. The coefficient is calculated
by integrating the normalized energy flux $S_z/S_z^i$ over the
slit value~\cite{Bet,Har}:
\begin{eqnarray}
T_{cw}=-\frac{\sqrt{\epsilon_1}}{4a \cos{\theta}}\int_{-a}^{a}
{\lim_{z\rightarrow{0^-}}}[(E_{x}H_{y}^*+E_{x}^*H_{y})]dx,
\end{eqnarray}
where $S_z^i$ is the energy fluxe of the the incident wave of unit
amplitude; $S_z$ is the transmitted flux. The first objective of
our computer analysis was to check the consistency of the results
by comparing the transmission coefficients calculated in the
studies~\cite{Bet,Har} with those obtained by our computations. We
have computed the coefficient $T_{cw}=T_{cw}(\lambda,a,b)$ for
different slit widths $2a$ and a variety of screen thicknesses b.
The results are presented in Fig. 2 for the wavelength
$\lambda$=500 nm. We notice that the transmission resonances of
$\lambda/2$ periodicity~\cite{Bet,Har} are reproduced.
Furthermore, the resonance positions and the peak heights
($T_{cw}{\approx}{\lambda}/{2\pi}a$~\cite{Bet}, at the resonances)
are in agreement with the results~\cite{Bet,Har}. Notice, that
$T_{cw}{\approx}{\lambda}/{2\pi}a > 1$ in the case of $a <
{\lambda}/{2\pi}$. The dispersion $T_{cw}=T_{cw}(\lambda)$ for the
given values of $a$ and $b$ is shown in Fig. 3. We notice that the
the amplitude of the short-wavelength waves is practically
unchanged, while the amplitude of the long-wavelength components
increases. Thus, effectively, the aperture "cuts off" the
short-wavelength FFT-components of the wave packet (6).

Owing to the "cut-off frequency" effect, the Fourier spectrum of
the wave-packet narrows under transmission through the aperture.
According to the Fourier analysis, the decrease of the spectral
width of the wave packet leads to increase of the pulse duration
and to modification of the temporal resolution associated with the
incident pulse. The capabilities and limits of simultaneous
spatial and temporal resolutions of the NSOM-pulse system are not
known. From a theoretical point of view, the most important
questions are the degree of collimating, the duration of the
ultra-short pulse past the sub-wavelength aperture and the rate of
spatial and temporal broadening of the pulse farther from the
aperture. To address these questions, the near-field amplitude
distributions of fs and sub-fs pulses passed through a nm-size
slit-type aperture was computed. The amplitude distribution
$|{\vec{E_x}}|$ of the diffracted wave-packet was computed for
different values of the incident-pulse duration $\tau$, central
wavelength $\lambda_0=2{\pi}c/\omega_{0}$, slit width $2a$ and
screen sickness $2b$. As an example, the distribution
$|{\vec{E_x}}|$ is shown at the three distances $z$ from the
screen: at the edge of the screen $(z=-0.1a)$, in the near-field
$(z=-a)$ and far-field $(z=-10a)$ zones of the diffraction (Figs.
4 and 5). Figures 4 and 5 show the amplitude distributions for the
cases of $\tau$ = 2 fs and 750 as, respectively. Notice, that the
value $2a=50$ nm is the minimum aperture size, which is generally
accepted for the practical near-field ($z\approx{-a}$) microscopy
in the visible spectral region
($\lambda\approx{500}$nm)~\cite{Bet}.

Analysis of Figs. 4(a) and 5(a) shows that the pulse is collimated
to exactly the aperture width at the edge of the screen. Hence,
the basic concept of NSOM remains valid for the ultra-short
pulses: the fs or sub-fs pulse passing through a nm-size aperture
can be used to provide a sub-wavelength (nm-scale) image,
super-resolution in space. We notice that in the case of Fig. 4a,
the duration ${\tau}'$ of the diffracted pulse at the screen edge
is practically the same as that of the incident pulse
(${\tau}'\approx{2}$fs). Thus, the temporal resolution associated
with the duration of the incident 2-fs pulse is practically
unchanged past the aperture. This demonstrates the possibility of
the simultaneous nm-scale resolution in space and the fs
resolution in time. In the case of the attosecond pulse
($\tau=750$as), the Fourier spectrum of the transmitted pulse
experiences sufficient narrowing. This leads to the increase of
the pulse duration in a few times (Fig. 5(a)). In the visible
spectral region, this effect imposes a limit on the simultaneous
spatial and temporal resolutions. It could be noted that the
increase of the pulse duration in the case of the attosecond
pulses can be reduced by decreasing of the central wavelength of
the incident wave-packet~\cite{Kuk1}. Further analysis of Figs.
4(a) and 5(a) indicates that the amplitude distribution of the
diffracted wave-packet is characterized by the formation of maxima
at the rims of the aperture. Notice, that the similar effect
exists in the case of the near-field diffraction of a continuous
wave~\cite{Nov}.

The results presented in Figs. 4(b) and 5(b) indicate the
possibilities and limits of the simultaneous spatial and temporal
resolutions at the distance $(z=-a)$. In the case of the 2-fs
incident pulse, the spatial resolution of the NSOM-pulse system is
approximately equal to the geometrical projection of the aperture
at the distance of half the aperture width (see, Fig. 4b). This
result demonstrates the practical possibility of the simultaneous
50-nm resolution in space and the fs resolution in time. The value
of 50 nm is the practical limit of the spatial resolution also for
NSOM the NSOM performing with a continuous wave~\cite{Bet}. The
spatial resolution of the NSOM-pulse system in the case of the
750-as incident pulse (see, Fig. 5(b)) is approximately two times
lower respect to the case of Fig. 4(b). This results indicates
that the spatial resolution of the NSOM-pulse system in the sub-fs
domain depends on the value of the incident-pulse duration. It can
be noted that spatial resolution of the SNOM-pulse system
performing with ps pulses is defined only by the aperture size and
independent of the wavelength~\cite{Mit1}. It is clear now that
the spatial resolution of the NSOM-pulse system is different for
each FFT $\omega$-component of the wave packet. The lowest spatial
resolution is achieved for the FFT $\omega$-component having the
lowest frequency $\omega_{LW}$ . The value
$\omega_{LW}=\omega_{0}-\Delta{\omega}/2$ depends on the central
frequency $\omega_0$ and spectral width $\Delta{\omega}$ of the
wave-packet. The width $\Delta{\omega}\approx{1/\tau}$ increases
with decreasing of the duration $\tau$. Therefore, the spatial
resolution of the NSOM-pulse system increases with the increase of
the values of $\tau$ and $\omega_0$.

In the far-field region $(z=-10a)$, we notice the considerable
temporal and spatial broadenings and the wave-shape changes in the
cases of the 2-fs and 750-as incident pulses (Figs. 4(c) and
5(c)). We also notice the existence of the negative time delay of
the diffracted pulse respect to the point $t=0$. The time
advancement increases with the increase of the distance $z$ (see,
Figs. 4(a-c) and 5(a-c)). The time-advancement effect is in
agreement with the superluminar behaviour of the light pulses
described in the study~\cite{Wyn}.

In the present model of the NSOM-pulse system we used the standard
boundary conditions based on the assumption of the perfect
conductivity of the screen. The super-resolution capabilities of
the system in time and space is a consequence of the assumption.
The perfect conductivity is a good approximation in a situation
involving a thick metal screen of sufficient opacity for a fs
wave-packet having the central wavelength ($\lambda_{0}=500$nm) in
the visible spectral region. However, in the case of sub-fs pulses
the decrease of the metal conductivity with decreasing the
wavelength should be taken into account for the short-wavelength
components of the wave-packet. The conductivity dispersion should
lead to some decrease of the spatial and temporal resolutions of
the NSOM-pulse system based on the use of sub-fs pulses. It should
be noted that the presence of a microscopic sample (a molecule,
for example) placed in strong interaction with the NSOM aperture
modifies the boundary conditions. In the case of the strong
interaction, which takes place in the region $z<-0.1a$, the
application of the standard boundary conditions practically
impossible. Probably, in this region the problem can be overcome
by combination of a microscopic and macroscopic descriptions of
NSOM~\cite{Gir}, where the response function accounting for the
modification of the quantum mechanical behaviour of the molecule
is derived self-consistently through the solution of Dyson's
equation.

\section{Conclusion}

The near-field diffraction of fs and sub-fs light pulses by
nm-size slit-type apertures and its implication for near-field
scanning optical microscopy (NSOM) has been analyzed. The
amplitude distributions of the diffracted wave-packets having the
central wavelengths in the visible spectral region were have been
found by using the Neerhoff and Mur coupled integral equations,
which were solved numerically for each Fourier's component of the
wave-packet. In the case of fs pulses, the duration and transverse
dimensions of the diffracted pulse remain practically the same as
that of the input pulse. This demonstrates feasibility of the NSOM
in which a fs pulse is used to provide the fs temporal resolution
together with nm-scale spatial resolution. In the sub-fs domain,
the Fourier spectrum of the transmitted pulse experiences a
considerable narrowing that leads to the increase of the pulse
duration in a few times. This imposes a limit on the simultaneous
resolutions in time and space. The results demonstrate the
capabilities and limits of simultaneous spatial and temporal
resolutions of the NSOM based on the use of the ultra-short light
pulses.

\begin{acknowledgments}
This study was supported by the Fifth Framework of the European
Commission (Financial support from the EC for shared-cost RTD
actions: research and technological development projects,
demonstration projects and combined projects. Contract
NG6RD-CT-2001-00602). The authors thank the Computing Services
Centre, Faculty of Science, University of Pecs, for providing
computational resources.
\end{acknowledgments}
%

% figures follow here
\newpage
\begin{figure}
\includegraphics[keepaspectratio,width=15cm]{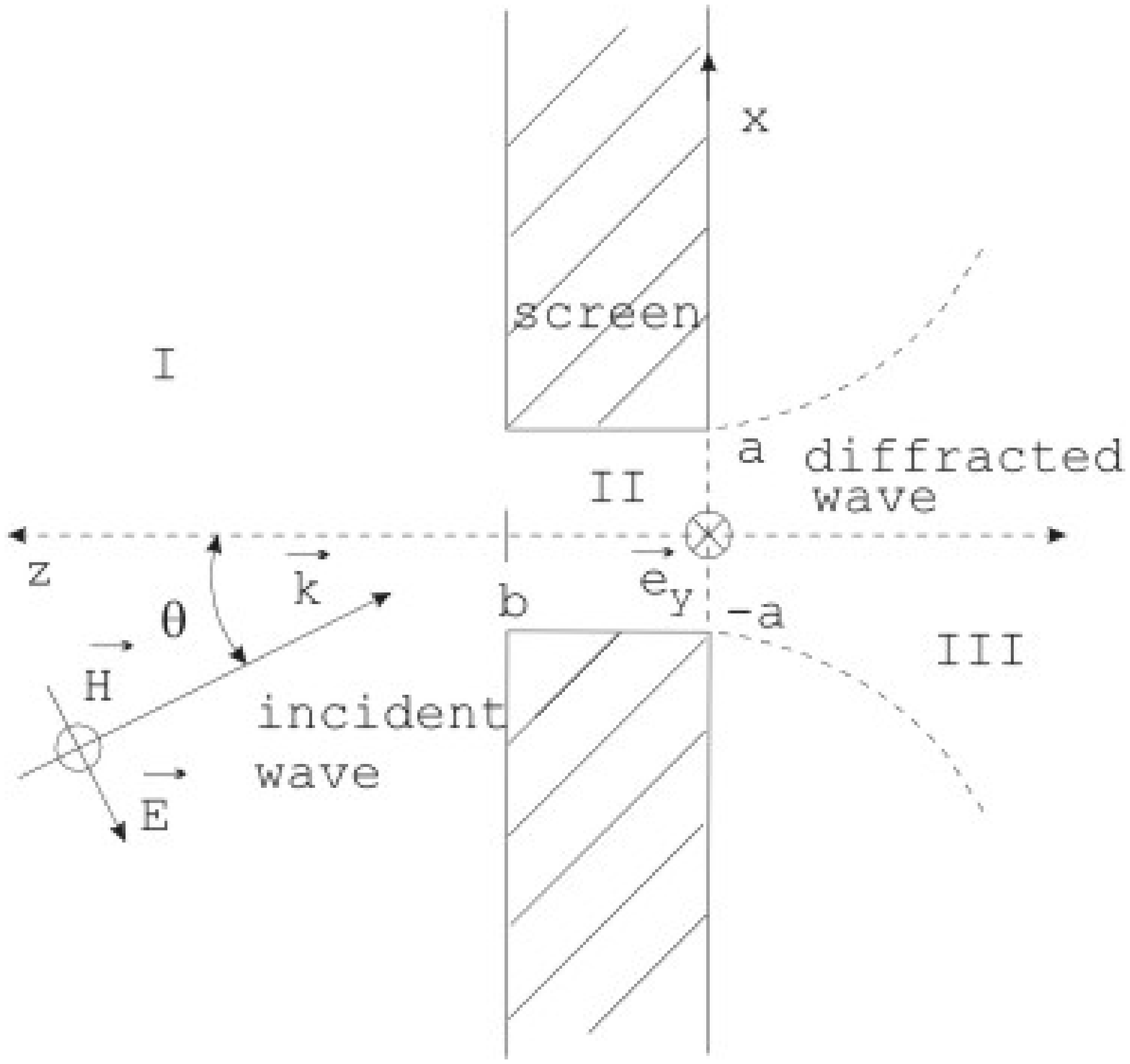}
\caption{\label{fig:epsart} Schematic diagram of NSOM.}
\end{figure}
\newpage
\begin{figure}
\includegraphics[keepaspectratio,width=15cm]{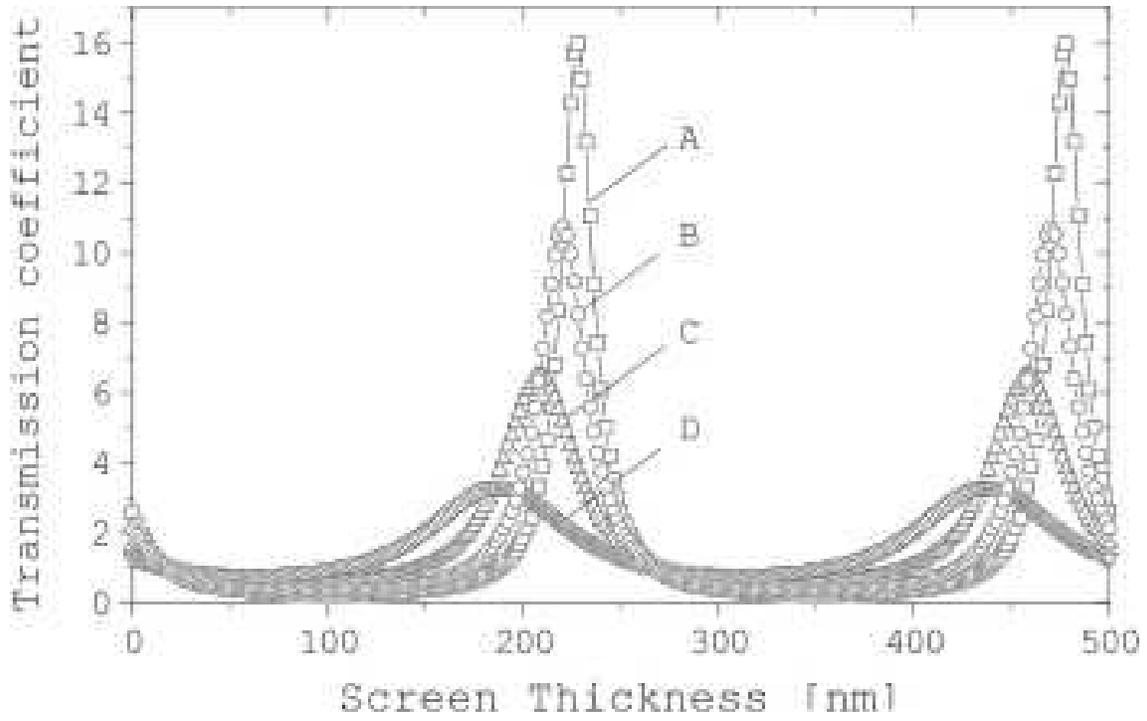}
\caption{\label{fig:epsart} Transmission
 coefficients for a continuous wave having $\lambda=500$nm as a
 function of screen thickness $b$ for the different aperture widths
 $2a$: $A$ - 10 nm, $B$ - 15 nm, $C$ - 25 nm and $D$ - 50 nm.}
\end{figure}
\newpage
\begin{figure}
\includegraphics[keepaspectratio,width=15cm]{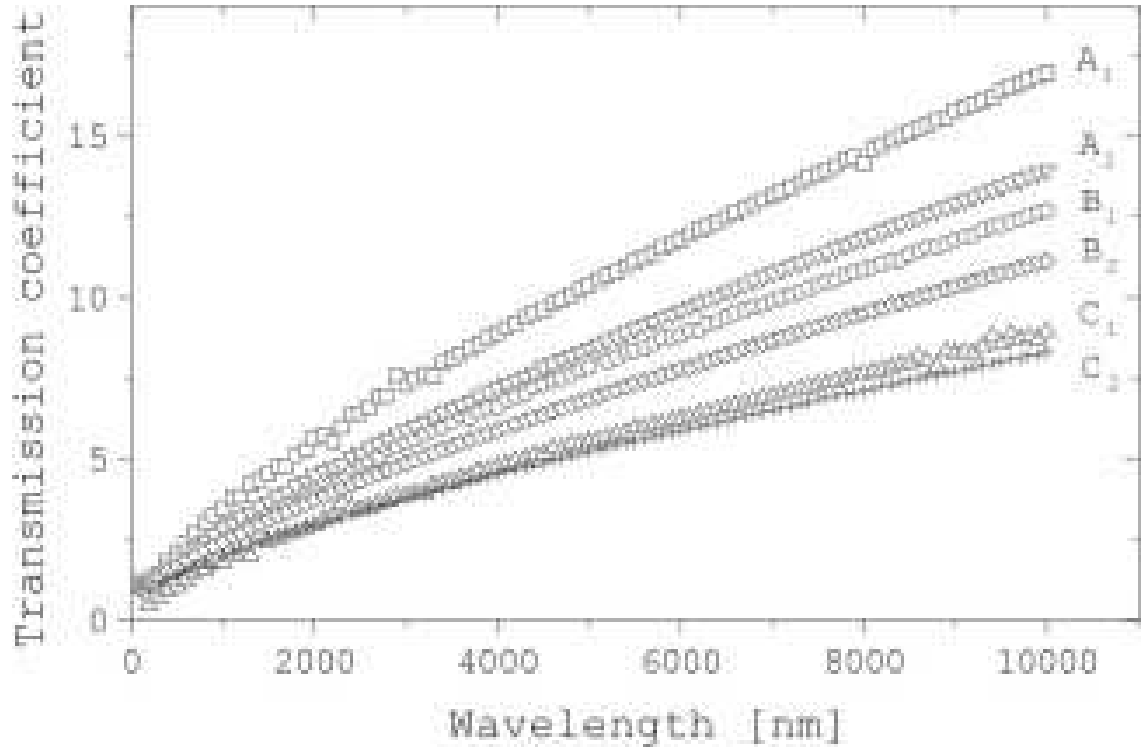}
\caption{\label{fig:epsart} The dependence
$T_{cw}=T_{cw}(\lambda)$ for the given values of $a$ and $b$. For
$b=0$, the values $2a$: $A_1$ - 10 nm, $B_1$ - 15 nm and $C_1$ -
25 nm. For $b=5$nm, the values $2a$: $A_2$ - 10 nm, $B_2$ - 15 nm
and $C_2$ - 25 nm.}
\end{figure}
\newpage
\begin{figure}
\includegraphics[keepaspectratio,width=12cm]{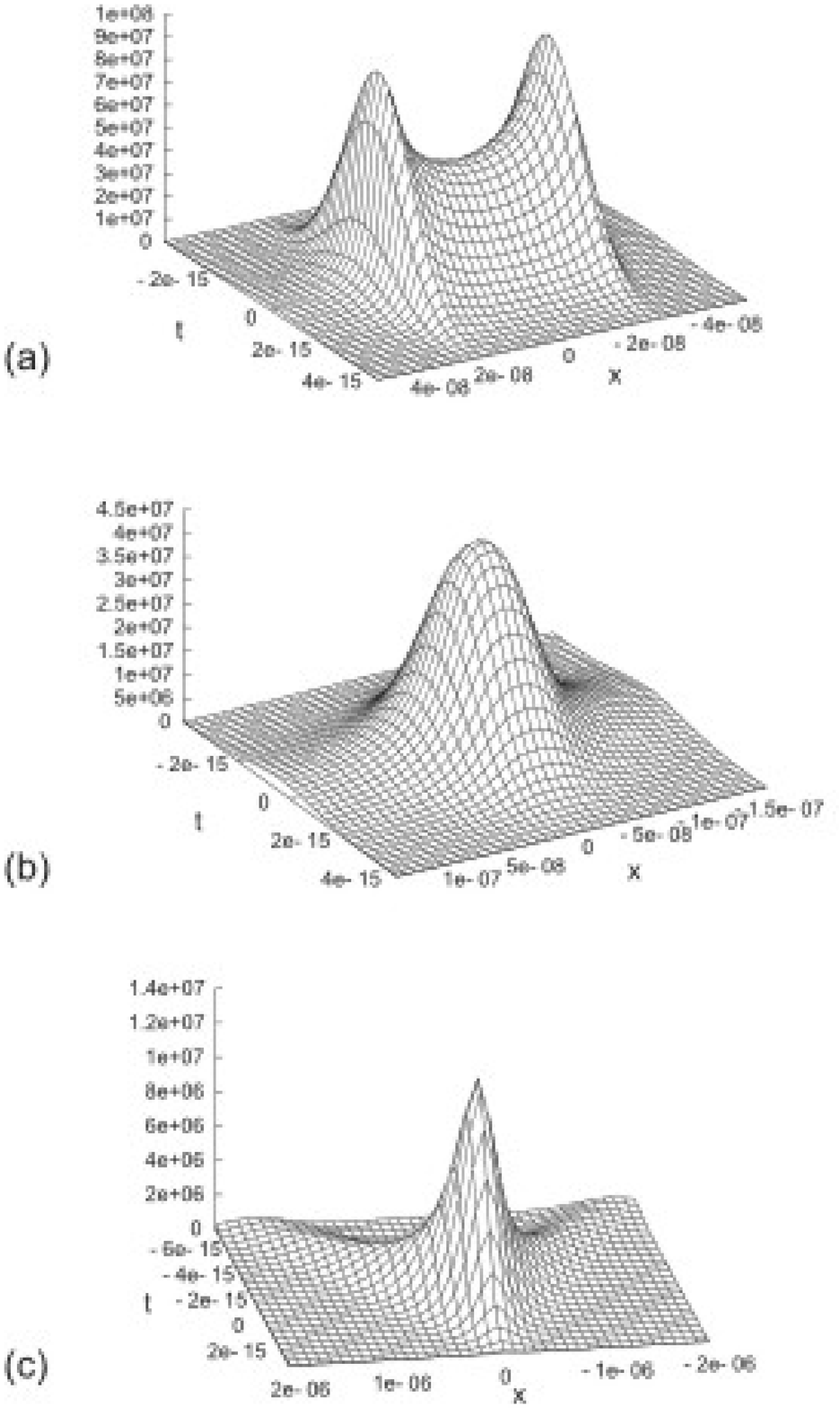}
\caption{\label{fig:epsart} The amplitude distribution
$|{\vec{E_x}}|$, in the arbitrary units, at the three distances
$z$ from the screen: (a) - $(z=-0.1a)$, (b) - $(z=-a)$ and (c) -
$(z=-10a)$. Here, the incident-pulse duration $\tau=2$fs, the
aperture width $2a=50$nm and thickness $b=25$nm, the packet
central wavelength $\lambda_{0}=500$nm and the amplitude
$U(x,z=b)=1$. The time $t$ and coordinate $x$ are in the second
and meter units, respectively.}
\end{figure}
\newpage
\begin{figure}
\includegraphics[keepaspectratio,width=12cm]{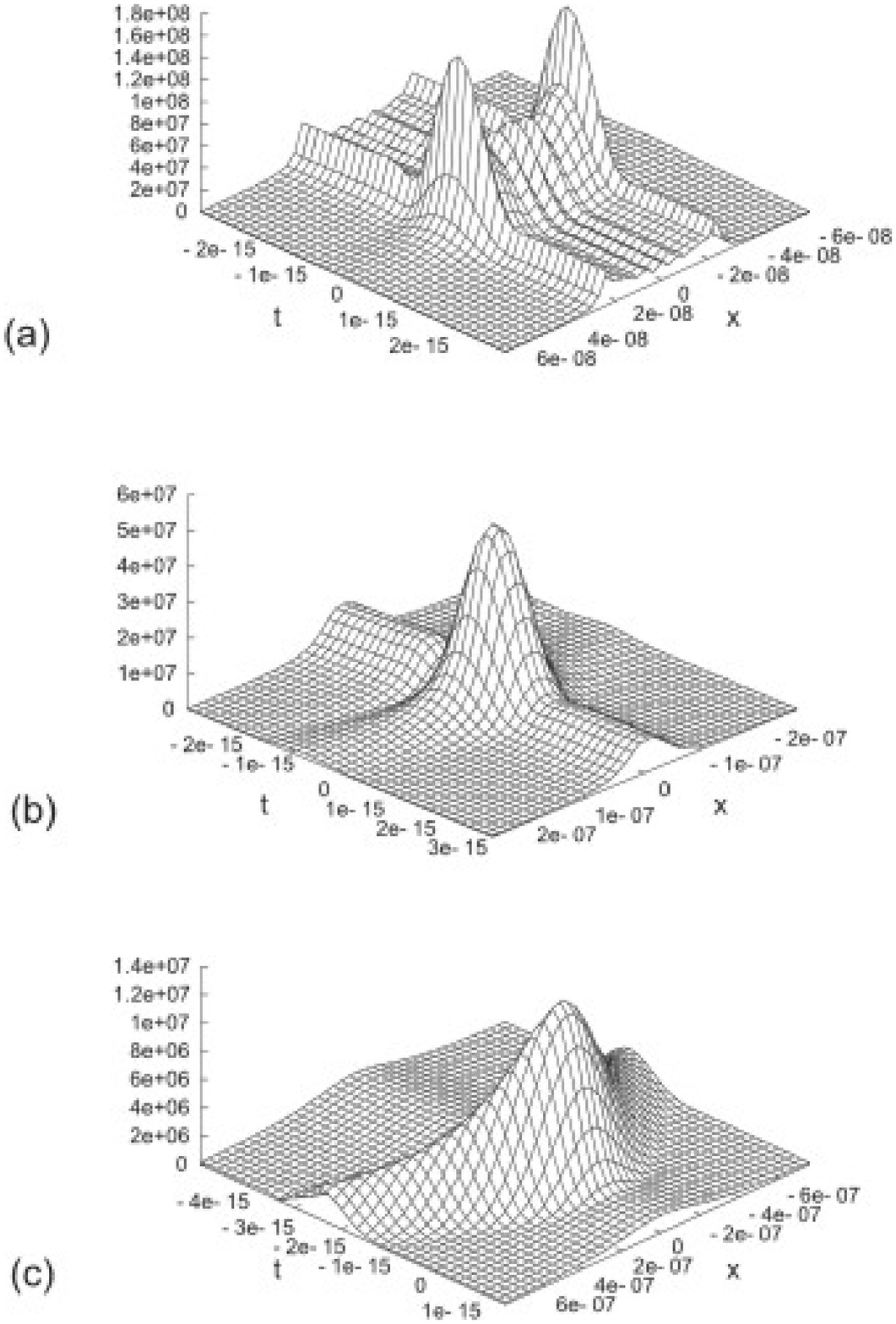}
\caption{\label{fig:epsart} The amplitude distribution
$|{\vec{E_x}}|$, in the arbitrary units, at the three distances
$z$ from the screen: (a) - $(z=-0.1a)$, (b) - $(z=-a)$ and (c) -
$(z=-10a)$. Here, the incident-pulse duration $\tau=750$as, the
aperture width $2a=50$nm and thickness $b=25$nm, the packet
central wavelength $\lambda_{0}=500$nm and the amplitude
$U(x,z=b)=1$. The time $t$ and coordinate $x$ are in the second
and meter units, respectively.}
\end{figure}
\end{document}